# Rashba spin-orbit interaction induced modulation of magnetic anisotropy

Megha Vagadia, Jaya Prakash Sahoo, Ankit Kumar, Suman Sardar, Tejas Tank and D.S. Rana

*Department of Physics, Indian Institute of Science Education and Research Bhopal, M.P. 462066, India*

**Abstract**

In past few decades, Rashba spin-orbit coupling (SOC) has been successfully employed for the emergence of exotic phenomena at the quantum oxide interfaces. In these systems, the combined effect of charge transfer, broken symmetries and SOC yields intriguing interfacial magnetism and transport properties. Here, we provide an insight to control and tune interfacial phenomena in $CaMnO_3/CaIrO_3$ based 3d-5d oxide heterostructures by the charge transfer driven Rashba SOC. Anomalous Hall effect in these canted antiferromagnetic heterostructures originates from the intrinsic contribution associated with the topology of the electronic band structure and it is mostly confined to the interface. Rashba SOC reconstructs the Berry curvature and enhances the anomalous Hall conductivity by two orders of magnitude. From the anisotropy magnetoresistance measurements we demonstrate that Rashba SOC is instrumental in tailoring magnetic anisotropy where magnetization easy-axis rotates from the out-of-plane direction to the in-plane direction. The ability to tune Rashba SOC and resulting competing magnetic anisotropy provides a route to manipulate electronic band structure for the origin of non-trivial spin texture useful for spin-orbitronics applications.

In complex quantum materials, spontaneous symmetry breaking is at the core of emergent new phases that are different than their microscopic picture. Remarkable phenomena like superconductivity antiferromagentism, ferroelectricity, and many more in condensed matter systems are results of broken symmetries [1, 2]. In the recent times, breaking of space inversion symmetry in the spin-orbit coupled systems is being explored for the efficient spin-to-charge conversion – a feature that is driving the development of the emerging field of spin-orbitronics [3-5]. Here, spatial asymmetry at the 2D interfaces induces the perpendicular built-in-electric field which in the presence of spin-orbit interaction lifts the degeneracy of the electronic band structure with the momentum dependent spin-split subbands. Charge carriers passing through this electric field experience an effective magnetic field that gives rise to momentum dependent Zeeman energy – also known as the Rashba spin-orbit coupling [6, 7]. The associated Rashba Hamiltonian is written as : $H_R = \alpha_R(k \times \sigma) \cdot z$, where $\sigma$ is the vector of the Pauli spin matrices, z is unit vector perpendicular to the interface, $k$ is the momentum and $\alpha_R$ is the Rashba coefficient which varies linearly with the electric field strength and SOC. The effective Rashba magnetic field can be derived from $B_R = 2\alpha_R k_F / g\mu_B$, here $k_F$ represent the Fermi wavevector and $g$ being the $g$-factor of the carriers in the conduction channel.

In spin-orbit coupled systems, the presence of electric field perturbs the electron's trajectory based on the direction of the angular momentum. This anomalous contribution to the velocity is induced by the Lorentz force produced by the Berry curvature, $\Omega(p) \propto \alpha_R \nabla_p \times (\sigma \times p)$ which depends on the intrinsic properties of occupied electronic states [8]. Non-vanishing Berry curvature in the systems with the broken time reversal symmetry gives rise to off-diagonal conductivity which contributes to the intrinsic anomalous Hall effect (AHE). Various intriguing physical properties of Rashba materials are explained in context of Berry curvature and associated Berry phase. The resultant modification in the velocity due to the Rashba effect can be deduced according to $V_a = \partial_p \hat{H}_R = -(\alpha_R/\hbar) z \times \sigma$ [3].

In 3d-5d transition metal based heterostructure, SOC has been key in realizing varieties of exotic phenomena at the interface such as spin-momentum locking, novel magnetic phases, intrinsic AHE, topologically non-trivial spin textures, etc. [9-12]. In addition, charge transfer in oxide heterostructures builts voltage potential across the interface that activates the Rashba effect [13]. The entanglement of the spin and orbital degrees of freedom in these heterostructures can be executed through Rashba effect in controlling and manipulating the spin currents. The recent demonstration of a giant spin-to-charge conversion efficiency in $SrTiO_3$ – based 2D electron gas system has paved potential route to implement Rashba physics for the oxide spin-orbitronics [4]. Further, SOC along with the crystal symmetry determines the magnetic anisotropy of the magnetic materials. Hence, many interfacial spin based phenomena and magnetic textures can be controlled and tuned intrinsically by SOC. For example, rotation of magnetization easy axis in $La_{2/3}Sr_{1/3}MnO_3/SrIrO_3$ heterostructures has been attributed to the strong SOC mediated through the emergence of a novel spin-orbit state in paramagnetic $SrIrO_3$ [14]. The competition between magnetic anisotropies can lead to the non-collinear spin texture with the topological band structure. Extrinsically, modulation in the magnetic anisotropy can be driven by the applied electric field. Despite promising transport phenomena induced by Rashba SOC in the presence of effective bias voltage at the oxide interfaces [13, 15-16], an important challenge is to enhance and control Rashba SOC effect in the absence of the bias voltage for the energy efficient oxide spin-orbitronics application.

Strength of Rashba SOC in oxide heterostructures can be tuned by controlling the fraction of charge transfer across the interface. In this work, we present the results of transport studies on $[(CaMnO_3)_x/(CaIrO_3)_y]_z$ (x, y = number of unit cells (u.c.)/period; z = repetitions) heterostructures; having charge transfer that varies with the constituent layer thickness. These heterostructures with canted antiferromagnetic structure exhibits Rashba SOC sensitive anomalous Hall effect arising from the Berry curvature; with the Rashba SOC controlled sign and magnitude of the anomalous Hall resistivity. As the strength of the Rashba field increases, the magnetic anisotropy of heterostructures change via the rotation of the magnetization easy axis from the out-of-plane to the in-plane direction and also show the sign reversal in the magnitude of anisotropic magnetoresistance. Present results provide an approach to tune AHE, magnetic anisotropy and electronic band structure in the quantum oxide heterostructures via the Rashba SOC in the absence of external bias voltage.

**Results and Discussion**

$[(CaMnO_3)_x/(CaIrO_3)_y]_z$ heterostructures studied here were fabricated via pulsed-interval epitaxy using a RHEED-assisted pulsed laser deposition technique onto $SrTiO_3$ *(100)* substrates. The X-ray reflectivity data and HAADF-STEM imaging suggested near atomically sharp interfaces for $[(CaMnO_3)_x/(CaIrO_3)_y]_z$ superlattices [17]. To estimate the element specific charge transfer in these heterostructures, we employed the x-ray absorption spectroscopy (XAS) at the Mn and Ir L-edges. Results of this study reported in [17] confirm transfer of charge from the Ir with Mn at the interface. In addition, a nearly constant shift in Ir edge for superlattices with varying thickness suggested that $CaIrO_3$ tends to lose only a constant fraction of its charge. Further, the charge transfer depends on the number of both $CaIrO_3$ and $CaMnO_3$ layers (and available carriers). In this context, three superlattices labelled as $(MIxy)_z$, (where M and I refer to $CaMnO_3$ and $CaIrO_3$ layers, respectively) - $(MI84)_5$, $(MI58)_4$ and $(MI22)_8$ with successive decrease in the charge transfer were chosen to understand the influence of interfacial Rashba SOC in controlling the exotic transport phenomena in oxide heterostructures. Magnetic and electrical properties of $[(CaMnO_3)_x/(CaIrO_3)_y]_z$ superlattices are found to depend strongly on the individual $CaMnO_3$ and $CaIrO_3$ layers thickness. These superlattices exhibit canted antiferromagnetic state where the magnetic transition temperature and saturation magnetization decrease with increasing period. Detailed results are discussed in [17] and supplementary information of this article.

**Magneto-transport behaviour of heterostructures**

Figure 1 (a) shows the magnetoresistance $[MR\ (H) = \frac{R(H)}{R(0)} - 1]$ measured at various temperatures for $(MI84)_5$ heterostructure. For H ∥ a, MR was measured with the current applied along the [001] direction. At 50 K, a cusp like feature in the low field region; with definite butterfly hysteresis appears which broadens with decreasing temperature. Hysteresis in MR indicates the presence of ferromagnetic interactions in the superlattices. Interestingly, below 30 K, in the low magnetic field region, the hysteresis becomes non-monotonic (Supplementary Fig. 1) and for 20 K again MR with the butterfly hysteresis is observed but with the opposite peak polarity (inset of figure 1(a)). This indicates the deviation in the magnetization easy axis from the out-of-plane [100] direction to the in-plane [010] direction. The results of the anisotropic magnetoresistance measurements further confirm the change in the magnetization easy axis in these heterostructures (discussed later). As seen from the comparison in figure 1 (b), at 10 K, all the three superlattices exhibit this cusp like feature with hysteresis. Detailed temperature dependence of MR (H) for all three superlattices is presented in Supplementary Fig. 1. With the decrease in charge transfer from $(MI84)_5$ to $(MI22)_8$, the presence of the hysteresis is moved towards low temperature with reduced width. Also, non-monotonic behaviour and peak polarity reversal is absent in $(MI58)_4$ and $(MI22)_8$ superlattices.

For the 2D-system, SOC is associated with weak antilocalization (WAL) [18, 19]. In this case the destructive interference of the backscattered electrons results in the cusp like feature at small magnetic field with positive MR. On the other hand, negative MR in the high magnetic field region originates from the weak localization (WL) effect due to the constructive back scattering of electrons [20, 21]. Similar observation is reported for thin metallic films and semiconductor as well as 3d-5d based oxide heterostructures [18, 22, 23, 24, 25]. The crossover from the positive to negative MR with increasing magnetic field suggests that these heterostructures have simultaneous presence of SOC and weak localization (WL) [25]. In such systems, Rashba SOC can be calculated according to the quantum correction theory of magneto-conductance associated with the WAL [26]. In order to deduce charge transfer induced interfacial Rashba SOC in $CaMnO_3/CaIrO_3$ heterostructures, we plotted the normalized magneto-conductance $\frac{\Delta G(H)}{G_0}$ $[\Delta\ G(H) = G(H) - G(0)]$ for all three superlattices at 10 K (figure 1 (c)). The coexistence of SOC induced WAL and WL results in the minima in the magnetoconductance. The magnetic field associated with this minimum ($B_{min}$) is found roughly proportional to the strength of SOC [27]. As expected, with the increase in the amount of the charge transfer from $(MI22)_8$, $(MI58)_4$ to (MI84)$_5$ the cusp feature becomes more prominent and the associated B$_{min}$ also increases; clearly indicating increase in the Rashba SOC strength. A similar enhancement of the cusp feature has been observed in the LaAlO$_3$/SrTiO$_3$ heterostructures under the application of large electric field of 100 V [28, 29].

Quantum corrected magnetoconductance of 2D system with SOC and WL at low perpendicular fields can be expressed by the Hikami-Larkin-Nagaoka (HLN) equation [26] as:

$$\frac{\Delta G\ (B)}{G_0} = -\psi\left(\frac{1}{2}+\frac{B_e}{B}\right) + \frac{3}{2}\psi\left(\frac{1}{2}+\frac{B_i+B_{so}}{B}\right) - \frac{1}{2}\psi\left(\frac{1}{2}+\frac{B_i}{B}\right) - ln\left(\frac{B_i+B_{so}}{B_e}\right) - \frac{1}{2}ln\left(\frac{B_i+B_{so}}{B_i}\right)$$

Here, $\psi$ is digamma function and $G_0$ is the universal conductance constant; $1.2 \times 10^{-5}$ S. $B_e$, $B_i$ and $B_{so}$ are effective fields of elastic, inelastic and SOC induced scattering terms, respectively. The HLN fit of magnetoconductance data at 10 K shown by the solid lines in figure 1 (c) indicates the good agreement of data with the theoretical model. We further determine the Rashba SOC coefficient from the $\alpha = \frac{(e\hbar^3 B_{so})^{1/2}}{m^*}$ relation where m* = 3m$_e$ is the effective mass of electron [30, 31]. Here, it should be noted that in the Ir-based oxide systems, the effective mass of the electron is found to exhibit film thickness dependence due to the quantum confinement [32]. However, for IrO$_2$, the change in effective mass is found to be only 10-20 % for the increase of film thickness from 3 to 9 monolayers. Similarly, for many other oxide and non-oxide systems also electron effective mass is found to be weakly dependent on the film thickness [33, 34, 35, 36, 37, 38, 39]. Hence, in this study we have considered effective mass to be constant for all three superlattice. Thus extracted parameters of B$_{so}$ and α from the HLN fit of the magnetoconductance are presented in figure 1 (d) demonstrate the enhancement in the Rashba SOC with the increase in the fraction of the charge transfer across the interface in the CaMnO$_3$/CaIrO$_3$ heterostructures.

**Anomalous Hall Effect (AHE) and its analysis**

Figures 2 (a) & (b) show the Hall resistivity ($\rho_{xy}$) as a function of magnetic field for (MI84)$_5$ and (MI58)$_4$ superlattices at various temperatures. Below 70 K, both the heterostructures exhibit anomalous Hall effect (AHE) with well saturated hysteresis. Recent reports suggest that 3d-5d based heterostructures with the non-collinear AFM structure can exhibit AHE [40, 41]. In materials having broken time reversal symmetry in the presence of strong SOC, intrinsic or extrinsic origin of AHE can be deduced from the relationship between anomalous Hall resistivity $\rho_{xy}^{AHE}$ and longitudinal resistivity $\rho_{xx}$ [10, 42]. A scattering rate independent $\rho_{xy}^{AHE}$ originating from the intrinsic mechanism varies quadratically to the longitudinal resistivity $\rho_{xx}$ (i.e. $\rho_{xy}^{AHE} \propto \rho_{xx}^2$) whereas extrinsic AHE arising due to the skew scattering varies linearly to $\rho_{xx}$ (i.e. $\rho_{xy}^{AHE} \propto \rho_{xx}$) [43-45]. The intrinsic AHE is associated to the topology of the electronic band structure via the Berry phase [42, 46].

To deduce the origin of AHE in CaMnO$_3$/CaIrO$_3$ heterostructures, in figure 2 (c), we plot $\rho_{xy}^{AHE}(T)$ as a function of $\rho_{xx}^2(T)(\mu_0 H = 0)$ for (MI84)$_5$ and (MI58)$_4$. The value of $\rho_{xy}^{AHE}(T)$ was taken for the field H = 9 T where magnetization and Hall resistivity both are saturated. For (MI84)$_5$, a quadratic dependence between $\rho_{xy}^{AHE}(T)$ and $\rho_{xx}$ indicates the scattering independent intrinsic origin of AHE invoking the Berry phase, whereas for (MI58)$_4$ $\rho_{xy}^{AHE}(T)$ deviates from the quadratic dependence. Speculating the origin of AHE in extrinsic skew scattering for (MI58)$_4$, we linearly fitted $\rho_{xy}^{AHE}$ as a function of $\rho_{xx}$, however, with no satisfactory fit (not shown here). Further, from the scaling behaviour, the obtained value of $\sigma_{xx} \sim 150 - 180 \, \Omega^{-1} cm^{-1}$ puts (MI58)$_4$ into moderately dirty metal limit. Based on the theoretical approach utilizing Berry phase and Berry curvature for the spin-orbit coupled material, in moderately dirty metal region intrinsic scattering leads $\rho_{xy}^{AHE}$ independent of $\rho_{xx}$ however, it is not followed in the case of (MI58)$_4$ [9, 47- 48]. The 3d-5d heterostructures near the moderately dirty metal limit have been reported to show characteristic dissipationless AHE [].

To get further insight into the origin of AHE we measured angular dependence of Hall effect at 10 K (figure 2 d). For both the heterostructures, magnitude of AHE decreases as the magnetic field direction rotates in the yz plane with an angle θ away from the normal axis, indicating the intrinsic nature of AHE arising from the Berry phase and Berry curvature [49]. Similar behaviour is also reported for the systems such as hcp Co single crystals [50,]. At higher temperature the angle dependence of AHE decreases suggesting the strength of Berry curvature reduces in accordance to the temperature dependence of AHE (Supplementary Fig. 2).

**Rashba spin-orbit coupling induced enhancement and sign reversal of AHE**

To understand the effect of Rashba SOC on Hall effect measurement, in figure 3 we compare Hall resistivity at 30 K for (MI22)$_8$, (MI58)$_4$ and (MI84)$_5$ having successively increasing Rashba SOC field. It should be noted here that, for (MI22)$_8$, the Hall resistance below 30 K was beyond the measurable limit of the instrument and the magnitude of $\rho_{xy}^{AHE}$ is very small which decreases with increasing temperature; therefore, no significant AHE signal was measured above 40 K. For (MI22)$_8$, the AHE is positive with $\rho_{xy}^{AHE}$>0 for M>0. With increase in Rashba SOC for (MI58)$_4$ and (MI84)$_5$, the sign of AHE reverses and negative sense of AHE with $\rho_{xy}^{AHE}$<0 for M>0 is observed with increasing hysteresis coercivity. The magnitude of AHE also enhances systematically with increasing Rashba SOC strength; an overall two order of magnitude increases from (MI22)$_8$ to (MI84)$_5$. Results of AHE show that Rashba

SOC via built-in-electric field at interfaces can greatly influence the electronic band properties of these heterostructures.

The results of ab initio calculations and XAS measurements reported in the literature suggest that for CaMnO$_3$/CaIrO$_3$ interfaces, the Mn-O-Mn along *xy*-plane are ferromagnetic double exchange interaction, whereas Ir-O-Mn interactions are antiparallel super-exchange along z-direction [11]. Charge transfer across these interfaces can induce the spin polarization [$P_s = \frac{DOS_\uparrow(E_F) - DOS_\downarrow(E_F)}{DOS_\uparrow(E_F) + DOS_\downarrow(E_F)}$] due to larger DOS ($E_F$) resulting from the different effective masses in t$_{2g}$ and e$_g$ bands. This spin polarization and hence the AHE, can be tune via the fraction of charge transfer (x) across the interface. For low x, $P_s > 0$ results in positive sense of AHE while for higher x, $P_s < 0$ gives negative sense of AHE.

This can be also understood in terms of the relationship between the intrinsic AHE and the integration of Berry curvature Ω of all the occupied Bloch bands in magnetic materials. In 2D ferromagnetic materials, the motion of the electrons can be affected by the non-zero Berry curvature in the momentum space giving rise to intrinsic AHE. In that case, the Hall conductivity (σ$_{xy}$) can be calculated by integrating the Berry curvature Ω in first Brillouin Zone as [51, 42]:

$$\sigma_{xy} = -\frac{e^2}{2\pi h}\int_{BZ}\Omega(\vec{k})d^2\vec{k}$$

Here, e is the elementary charge, h is the Planck constant, and $\vec{k}$ is the momentum wavevector. Under the application of perpendicular applied magnetic field, the spontaneously ordered magnetic moments deviate the motion of electrons to the high (low) potential side and the intrinsic AHE with positive (negative) sign emerges for $\int \Omega < 0$ ($\int \Omega > 0$). Thus, sign reversal of AHE in CaMnO$_3$/CaIrO$_3$ interfaces suggests that Rashba SOC significantly influence the electronic band structure via the reconstruction of the Berry curvature and Berry phase.

Magnetization results on these heterostructures show the highest saturation moment for thinner superlattice (MI22) which decrease with the increase in the thickness [17]. In contrast to magnetization, the stronger AHE is observed for (MI84)$_5$ superlattice having lowest magnetization. As discussed earlier, the fraction of the charge transfer across the interface influence the magnetic super exchange interaction in addition to the spin polarization. The Hamiltonian that depicts the magnetic moment alignment is given as [52]:

$$\widetilde{H} = -t\cos\left(\frac{\beta}{2}\right)\sum_{<ij>}(a_{i\uparrow}^\dagger a_{j\uparrow} + h.c.)\sum_{<ij>}J_{ex}(\widehat{m_i}\cdot\widehat{m_j})$$

which simplifies into an energy equation $E = -4x|t|cos\left(\frac{\beta}{2}\right) + J_{ex}\cos(\beta)$ to be minimized, where $a^{\dagger}/a$ is the creation/annihilation operators, t is hopping integral, β is the angle deviating from ferromagnetic alignment, $J_{ex}$ is the superexchange coupling and *x* is the charge transfer fraction. Hence, the amount of the charge fraction x transferred across the interface decides the alignment of Mn magnetic moment in $CaMnO_3$ layers. For adequately high x, Mn and Ir moments are antiferromagnetically coupled while Mn are ferromagnetically coupled [9, 52]. But low x, results in canted antiferromagnetic ordering of Mn moments along *xy*-plane. Thus, $(MI84)_5$ has higher $Mn^{3+}$ and $Mn^{4+}$ ratio indicating presence of larger ferromagnetic exchange of Mn moments along *xy*-plane whereas the bulk part of $CaMnO_3$ where charge transfer decays exponentially, antiferromagnetic interactions are dominant. These results indicate that majority of AHE resides at the interface and within $CaMnO_3$ layer.

The observed enhancement in AHE can also be correlated to the strength of SOC. According to the first order approximation the AHE is proportional to the SOC [42] and therefore the successive increase in the magnitude of the AHE from $(MI22)_8$ to $(MI84)_5$ also further the role of Rashba SOC in altering interfacial magnetism, AHE and electronic band structure in $CaMnO_3/CaIrO_3$ heterostructures. Similar results on the role of interface induced Rashba SOC in modifying AHE have been reported for p-i-n junctions, magnetic topological insulator heterostructures etc. [13, 53]

**Anisotropic magnetoresistance (AMR) measurements and modulation of magnetic anisotropy**

To probe the magnetic anisotropy of $CaMnO_3/CaIrO_3$ heterostructures, anisotropic magnetoresistance of all three superlattices was measured in detailed temperature and magnetic field range. As presented in the schematic (Inset of figure 4 (a)) the applied magnetic field was rotated in the yz plane with respect to the superlattice and AMR was calculated as:

$$\text{AMR (\%)} = \frac{\rho[B(\theta)] - \rho[B(\theta = 90°)]}{\rho[B(\theta = 90°)]} \times 100 \%$$

The current was passed along the pseudocubic [001] and θ denotes the angle between magnetic field and axis normal to the heterostructure. In figure 4 we present thus measured AMR for $(MI84)_5$ at 25 K with varying magnetic field strength. AMR exhibits mainly two-fold sinusoidal oscillations throughout the temperature and magnetic field range. AMR measured by clockwise and counter-clockwise rotation of sample shows the hysteresis (Supplementary Fig. S8).

The variation in the AMR magnitude for all three superlattices as a function of temperature presented in the supplementary information indicates the competition between temperature dependent spin-lattice and field-pseudospin couplings. As discussed in details in [17] and Supplementary Fig. 3, for thicker superlattice $(MI84)_5$ the AMR amplitude peaks around magnetic transition due to the dominant role of S-L coupling whereas for (MI58)$_4$ and (MI22)$_8$ the maximum AMR is obtained at the lower temperature i.e. below 10 K indicating a dominant role of field-pseudospin coupling in addition to the S-L coupling.

The most remarkable feature of this AMR is the $\pi/2$ phase shift in the peak position accompanied by the sign reversal. As we see in figure 4 (a) for low magnetic field, the minimum is observed around $\theta = 180°$ indicating easy axis is close to the out-of-plane direction, here AMR mostly is positive. In the intermediate magnetic field range, the minima gradually shift towards lower $\theta$ with the positive to negative crossover in the AMR sign. For the higher field of 7 T and above, maximum phase shift of $\pi/2$ with the minima around $\theta = 90°$ reveals that the easy axis is now close to the in-plane direction. For (MI84)$_5$, such modulation in the magnetic anisotropy is also observed as a function of temperature. The comparison of AMR polar plots in figure 4 (b) for T = 2 K and 40 K at H = 7 T depicts clear phase shift of $\pi/2$. Additional AMR data with the detailed comparison as a function of temperature and magnetic field can be seen in Supplementary Fig. 4. As evident from the figure 4 (c), for (MI84)$_5$ this modulation of magnetic anisotropy can be tuned systematically over the elaborated temperature and magnetic field range. The scattered data point represents the dominant AMR % i.e. dominance of AMR oscillation in the positive or negative region in varying temperature and magnetic field range; red symbols for positive AMR and blue for negative AMR. Inset of figure 4(c) suggest that for 2 K, AMR is negative throughout the range of magnetic field. Circular symbols in figure 4(c) shows the variation in AMR ratio with temperature for H = 9 T. A transition in the sign reversal starts at 14 K. Magnetic field dependence for 20 K and 30 K presented by square and diamond symbols, respectively, reveals that the transition moves towards lower magnetic field with increase in the temperature. A systematic shift in the phase reversal is seen from the data in Supplementary Fig. 4. For 50 K a complete crossover of easy axis close to in-plane direction with positive sign dominates for all the magnetic field range (inset of figure 4 (c)). A close inspection of AMR oscillations in range of 8-20 K for H = 9 T reveals the tailoring of magnetic anisotropy via the slow and gradual rotation of the magnetization easy axis which is initiated through the difference in the scattering intensities associated with [100] and [$\bar{1}$00] directions (Supplementary Fig. 5).

In Figure 5 we compare $\theta$- AMR measured at various temperatures for H = 9 T for all the three superlattices under this study. With the decrease in the strength of Rashba SOC from (MI84)$_5$ to (MI58)$_4$, the transition accommodating phase shift moves towards higher temperature for H = 9 T whereas no magnetic field induced such transition is observed throughout the temperature range of measurements for (MI58)$_4$ (Supplementary Fig. 6). For (MI22)$_8$, with lowest strength of Rashba SOC, this transition vanishes completely for the function of both temperature and magnetic field (Supplementary Fig. 7).

Present results clearly highlight the role of Rashba SOC in tailoring magnetic anisotropy of the oxide heterostructures.

Recently, owing to its technological importance in obtaining and tailoring non-collinear topological spin textures, in oxide heterostructures and thin films, efforts to modulate magnetization easy axis has been achieved by tuning strong SOC via the thickness control of $SrIrO_3$ layer, epitaxial strain, oxygen octahedra rotation, switching of noncollinear magnetic structure between AFM and FM states and by the spin flop transition [14, 54-57].

Extrinsic contribution such as presence of exchange bias field can also alter the magnetic anisotropy of the materials [58]. In this context, we also obtained exchange bias fields for presently studied heterostructures by measuring field cooled magnetization at 10 K. Data of these measurements shown in Supplementary Fig. 9 suggests the presence of exchange bias field of 35 Oe in $(MI84)_5$ due to co-existence of AFM and FM phases created by the exponentially decaying charge in the $CaMnO_3$ layer from the interface [17]. On the other hand, no exchange bias is observed for $(MI58)_4$ and $(MI22)_8$ superlattices. It should be noted that all the AMR measurements were carried out in the zero- field cooling protocol with erasing the magnetic history to minimize the exchange bias effect. In addition, $(MI58)_4$ superlattice with the moderate strength of Rashba SOC and no exchange bias also shows the temperature induced rotation of the magnetization easy axis. Further, exchange bias field effect is maximum at 10 K, at this temperature no magnetic field induced modulation is found in $(MI84)_5$ (Supplementary Fig. 10). Hence, the role of exchange bias in contributing the magnetic anisotropy can be understood not to dominate the Rashba SOC effect.

**Non-volatile switching of anisotropic resistive states**

Above AMR measurements suggest that tuning of the magnetocrystalline energy in canted antiferromagnetic structure can be utilized to design the non-volatile memory resistive switching where on and off states of the device can be tuned by the rotation of the magnetization easy axis. In the present study we propose the prototype of the functionality of such devices with different control parameter as the magnetic anisotropy is tuned by temperature, magnetic field, rotation angle and forwards and reverse sweeping of the AMR oscillations. Figures 6 show retention properties of the switching states controlled by magnetic field and angle (theta). For the switching in figure 6 (a), the on and off states were obtained by reorientation of the easy axis from the out-of-plane direction to in-plane direction by magnetic field of 1 T and 5 T, respectively. Figure 6 (b) demonstrates the retention property of the on and off states obtained (at T = 30 K and H = 7 T) by the reorientation of the spins by changing the angle theta between 25° and 107°, respectively.

In conclusion, our present study shows effective tuning of novel transport phenomena at 3d-5d interfaces by the Rashba spin-orbit coupling. In particular, Berry phase driven spin texture and magnetic anisotropy can be engineered by the pronounced spin-orbit coupling in the absence of an electric field. Switching between these anisotropy governed resistance states offers a prospective to design non-volatile multifunctional memory devices. This work also provides ground to formulate desirable control of Rashba SOC without applied bias voltage for operational spin-orbitronics application.

**Methods**

***Sample fabrication and structural characterization.*** $[(CaMnO_3)_x/(CaIrO_3)_y]_z$ (x, y = number of unit cells (u.c.)/period; z = repetitions) heterostructures were fabricated using pulsed laser deposition onto SrTiO$_3$ (STO) *(100)* substrates. The superlattices were deposited at a substrate temperature of 730°C and an oxygen partial pressure of 6 Pa followed by post-annealing at the same temperature and pressure for 30 minutes. The thickness of the superlattices was precisely controlled/monitored by *in-situ* reflection high-energy electron diffraction (RHEED). The high structural quality of the superlattices was examined in detail by performing room-temperature X-ray diffraction (XRD) $\theta$-$2\theta$ scans using a PANaylitcal X'pert Pro diffractometer. Out-of-the plane lattice parameters and strain states were studied by performing reciprocal space map (RSM) measurements, while the thicknesses of the superlattices were confirmed by X-ray reflectivity (XRR).

***Electrical characterization.*** Transport and magnetotransport measurements were done in a Quantum Design Physical Properties Measurement System (PPMS) equipped with rotator module. The resistivity measurements were performed in constant-current mode at 10 µA to avoid Joule heating. The electrical contacts were made with Al wire using a wire bonder (West Bond make) in the four-probe geometry for the resistivity measurements. To obtain the angular dependence of the resistivity, the sample orientation was varied with respect to the magnetic field at a fixed temperature and magnetic field. For the $\theta$-rotation, the magnetic field was applied perpendicular to the sample surface, and the current was applied along the *(hkl)* directions. The transvers DC resistance was measured as a function of magnetic field at various temperatures. The Hall resistance was corrected for misalignment of electrodes and magnetoresistance contribution by $R_{xy}(H) = \frac{(R_{xy}(+H) - R_{xy}(-H))}{2}$.

***Magnetic characterization.*** The magnetic behavior of the superlattices was investigated in a Quantum Design superconducting quantum interference device (MPMS-3). The temperature dependence of the magnetization was measured at various applied magnetic fields under the field-cooled protocol during the warming cycle. Magnetization isotherms were measured at various temperatures by sweeping the magnetic field between ± 7 T. In order to study the exchange bias effect in the superlattices, magnetization loops were obtained after cooling the samples in the presence of the applied magnetic field from the paramagnetic region to the temperature of the measurement.

**Acknowledgments**

D.S.R. thanks the Science and Engineering Research Board (SERB) Technology, New Delhi for financial support under the project no. CRG/2020/002338. M.V. acknowledges Department of Science and Technology (DST) New Delhi, India for the INSPIRE faculty award (DST/INSPIRE/04/2017/003059). J. S. acknowledges the DST-INSPIRE for Fellowship (DST/INSPIRE/03/2018/000699). Authors thank Dr. Ravi Shanker Singh (IISER Bhopal), Dr. Gulloo Lal Prajapati (Helmholtz-Zentrum Dresden-Rossendorf, Germany), Dr. T. Suraj (NUS, Singapore) and Dr. Ganesh Ji Omar (NUS, Singapore) for fruitful discussions.

Competing interests:
Authors declare no competing interests.

**Figures**

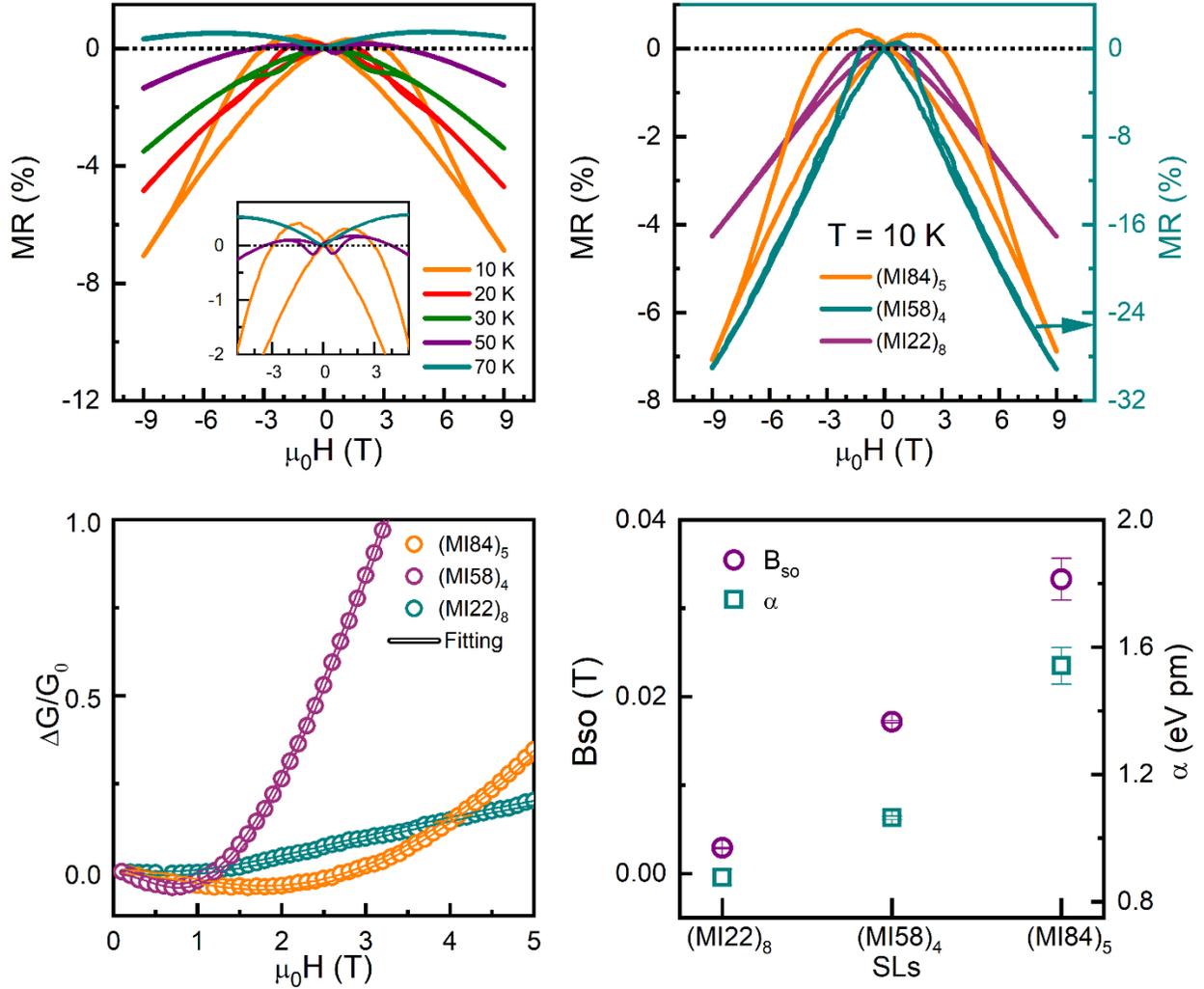

**Figure 1: Magnetoresistance measurements and Rashba spin-orbit coupling characteristics.** (a) Magnetoresistance $MR\ (H)\% = \left[\frac{R(H)}{R(0)} - 1\right] \times 100\ \%$ measured at various temperatures for (MI84)$_5$ heterostructure. For applied magnetic field along out-of-plane (i.e. H ∥ a) direction, MR was measured with the current applied along the in-plane (i.e. H ⊥ a) direction. (b) Comparison of MR at 10 K for (MIxy)$_z$ superlattices. Dashed lines in (a) and (b) denotes MR % = 0, indicating the crossover from weak anti-localization to weak localization. (c) Quantum corrected magnetoconductance ($\Delta G/G_0$) for (MIxy)$_z$ superlattices at T = 10 K with the fit according to the Hikami-Larkin-Nagaoka (HLN) equation. Open symbols show experimental data whereas the fit to HLN theory is represented by solid lines. (d) Variation in effective fields of elastic ($B_e$), spin-orbit coupling ($B_{so}$) scattering terms and the Rashba SOC coefficient (α) for (MIxy)$_Z$ superlattices with the varying fraction of charge transfer across the CaMnO$_3$/CaIrO$_3$ interfaces.

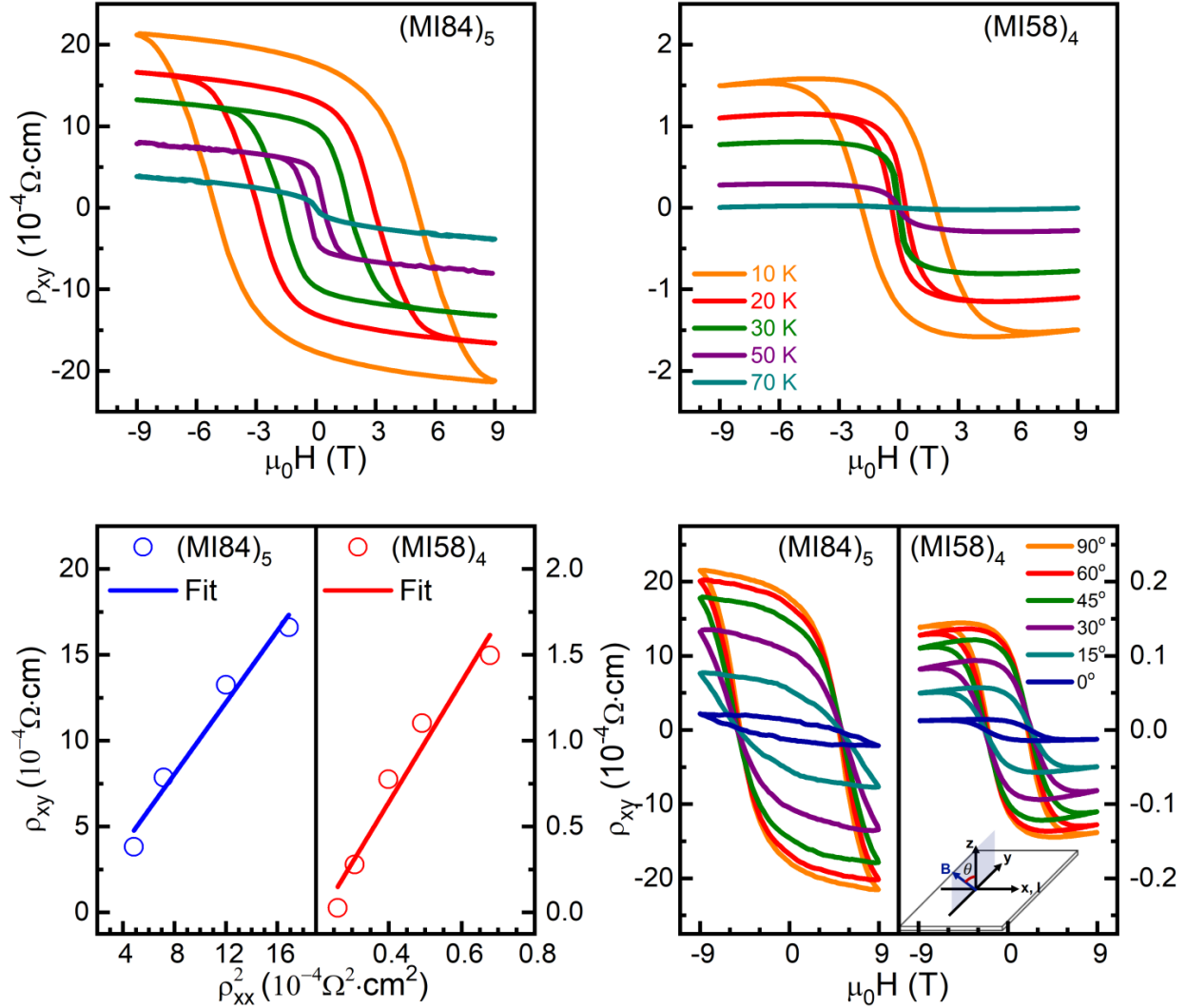

**Figure 2: Anomalous Hall effect measurements and analysis.** (a) Transverse Hall resistivity ($\rho_{xy}$) as a function of the perpendicular applied magnetic field measured at temperatures in 10 – 70 K range for (a) (MI84)$_5$ and (b) (MI58)$_4$ superlattices. (c) Plots of $\rho_{xy}$ vs. $\rho_{xx}^2$ for (MI84)$_5$ and (MI58)$_4$. Symbols represent experimental value of $\rho_{xx}(T)$ $at$ $\mu_0 H = 0$ and $\rho_{xy}^{AHE}(T)$ for the field H = 9 T where Hall resistivity is saturated. Solid line represents straight line fit to the experimental data. (d) Angular dependence of anomalous Hall effect at T = 30 K. Inset depicts the rotation geometry of the magnetic field with the current applied perpendicular to the magnetic field.

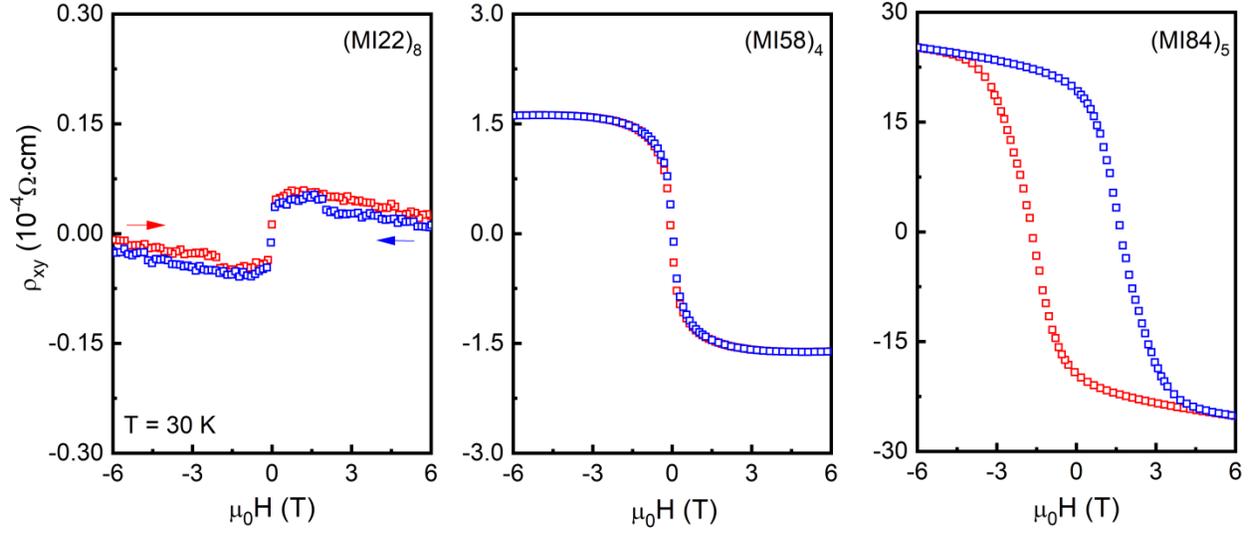

**Figure 3: Rashba spin orbit interaction induced alterations in anomalous Hall resistivity**. Transverse Hall resistivity for $(MI22)_8$, $(MI58)_4$ and $(MI84)_5$ measured at T = 30 K show sign reversal and enhancement in the anomalous Hall resistivity with successive increase in Rashba SOC.

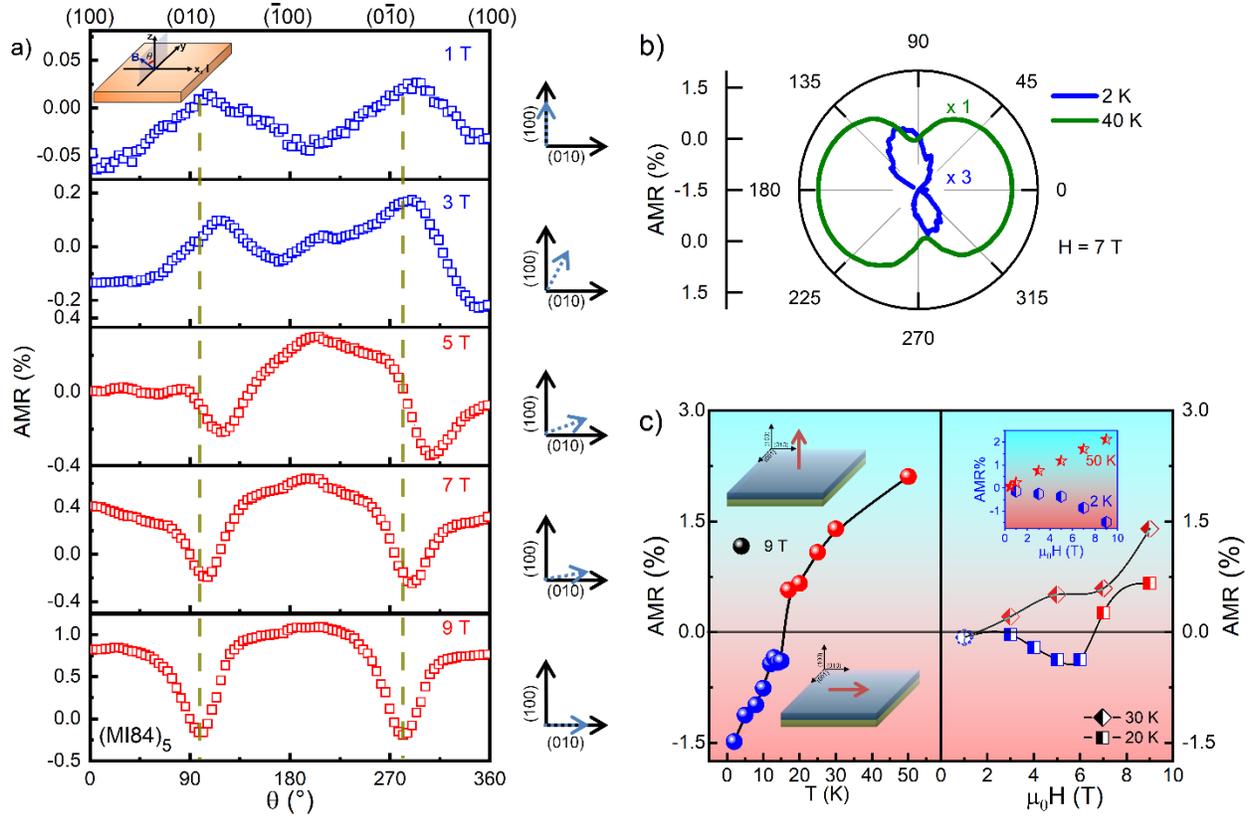

**Figure 4: Reversal in sign and phase of anisotropic magnetoresistance for (MI84)$_5$ superlattice.** (a) θ-AMR $[\rho(B(\theta)) - \rho(B(\theta = 90°))/\rho(B(\theta = 90°))]$ of (MI84)$_5$ at 25 K measured with increasing magnetic field strength. Schematic in the inset shows the applied magnetic field was rotated in the *yz*-plane with respect to the superlattice and the current was passed along the pseudocubic [001] direction. Dashed line guides the modulation of the anisotropy with magnetic field evident from the sign and phase change of AMR. (b) Polar plots comparing θ-AMR for T = 2 K and 40 K (H = 7 T) shows phase shift of π/2 indicating evolution of the magnetic anisotropy with temperature. (c) The scattered data point represents dominance of AMR oscillation in the positive (red symbols) or negative region (blue symbols). AMR data for 2 K and 50 K in the inset suggest absence of magnetic field mediated rotation of the easy axis. Temperature dependence of AMR for H = 9 T presented by circles shows the phase shift and sign reversal starts to appear at T = 14 K. With increase in the temperature, the modulation in the magnetization easy axis can be executed at lower magnetic field as represented by square and diamond symbols for 20 K and 30 K, respectively.

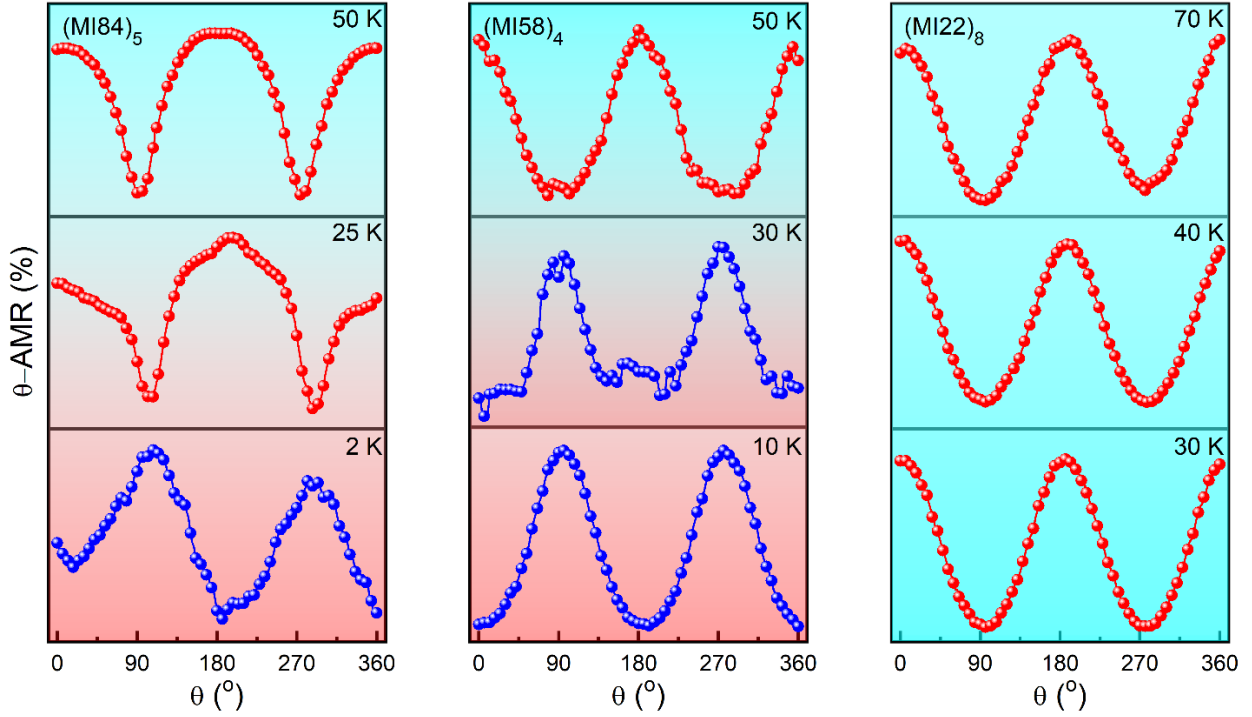

**Figure 5: Rashba SOC controlled modulation of magnetic anisotropy.** Temperature dependence of $\theta$-AMR ratio at H = 7 T for (MI84)$_5$, (MI58)$_4$ and (MI22)$_8$ presented in the decreasing order of the Rashba SOC strength. The transition accommodating the phase shift moves towards higher temperature with decrease in Rashba SOC and vanishes completely for (MI22)$_8$ with lowest strength of Rashba SOC.

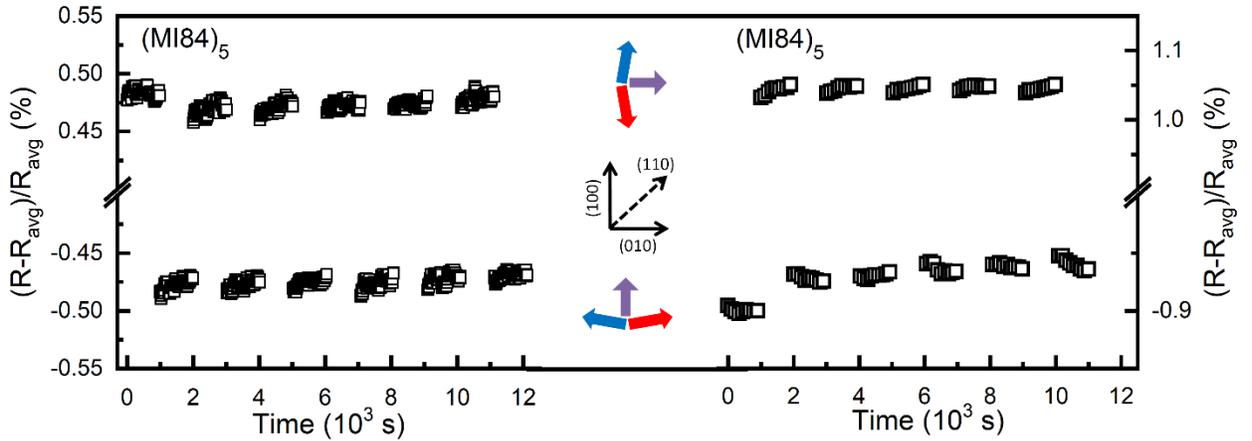

**Figure 6:** Retention property of the switching between On and Off states obtained by rotating the magnetization easy axis by magnetic field and angle θ.

Rashba spin-orbit interaction induced modulation of magnetic anisotropy

Megha Vagadia, Jaya Prakash Sahoo, Ankit Kumar, Suman Sardar, Tejas Tank and D.S. Rana

*Department of Physics, Indian Institute of Science Education and Research Bhopal, M.P. 462066, India*

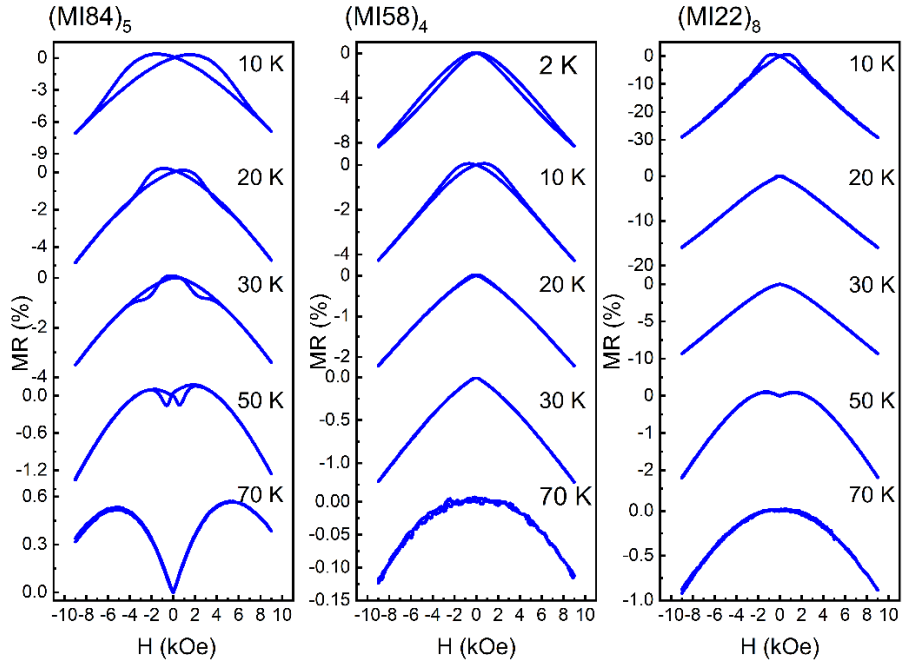

Figure S1: Magnetoresistance (MR) measured as a function of magnetic field at various temperature.

Figure S1 shows the magnetoresistance isotherms at various temperatures below magnetic transition for all three superlattices under study. For $(MI84)_5$, below 70 K, a butterfly shaped hysteresis opens with dip in the low magnetic field region. With lowering of temperature, a non-monotonic MR behaviour is observed at T = 30 K. Further, at 20 K, a butterfly shaped hysteresis appears with the peak in the low field region but opposite in the polarity than the dip at 50 K. Below, 20 K the width of the hysteresis increases significantly. This temperature induced non-monotonic evolution of the MR in the low field region suggest the deviation of the magnetization easy axis from the out-of-plane direction. For the superlattices $(MI58)_4$ and $(MI22)_8$ with decreasing amount of charge transfer, no such rotation of magnetization easy axis is observed whereas for these two superlattices hysteresis appear only below 10 K with the decreasing width with the decreasing charge transfer.

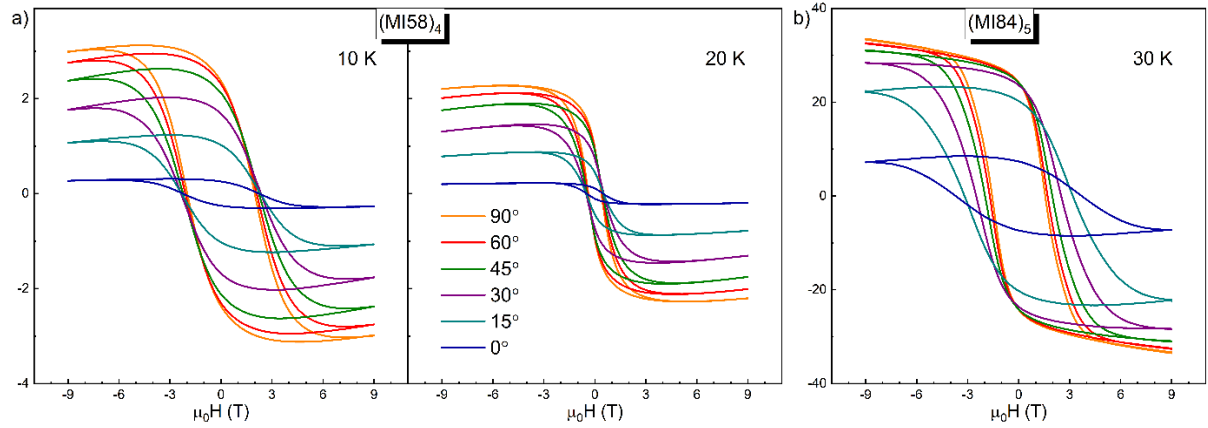

Figure S2: Angular dependence of anomalous Hall effect for (MI58)$_4$ and (MI84)$_5$.

Figure S2 depicts that both (MI84)$_5$ and (MI58)$_4$ superlattices exhibit strong angular dependence of AHE, supporting its origin in the intrinsic mechanism. With increase in the temperature the angle dependence of AHE decreases suggesting the strength of Berry curvature reduces in accordance to the temperature dependence of AHE.

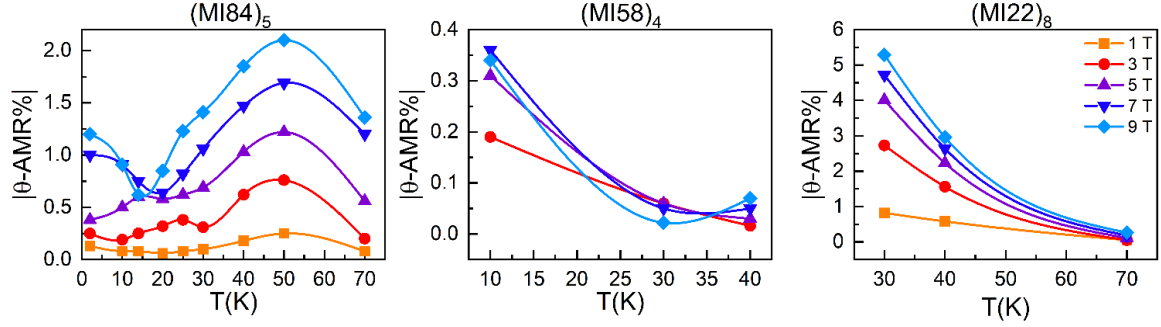

Figure S3: Variation in the AMR ratio as a function of temperature for (MIxy)z superlattices.

Variation in the AMR ratio as a function of temperature presented in figure S3 indicates the competition in temperature dependent spin-lattice and field-pseudospin couplings. The S-L coupling is strong in the vicinity of a magnetic transition and weakens on lowering the temperature. The in $CaIrO_3/SrTiO_3$ superlattices AMR is reported to scale with the strength of S-L coupling and thus peaks around the transition temperature. Contrary to this, in $CaIrO_3/CaMnO_3$ superlattices, the AMR peaks well below the magnetic transition; for superlattices $(MI58)_4$ and $(MI22)_8$ AMR amplitude monotonically increases with decreasing temperature showing the maximum AMR at much lower than the magnetic transition i.e. 10 K. For thick $(MI84)_5$ superlattice, however, the AMR peaks at the elevated temperature of ~ 50 K. This indicates a dominant role of field-pseudospin coupling in addition to the S-L coupling. As the out-of-plane axis rotates with respect to the field in *AMR* measurement, the field-lattice coupling rotates the orthorhombic distortion. At low temperatures, however, the stiffness of the lattice weakens the S-L coupling. This tilts the balance in favour of field-pseudospin coupling in the presence of large magnetic moments.

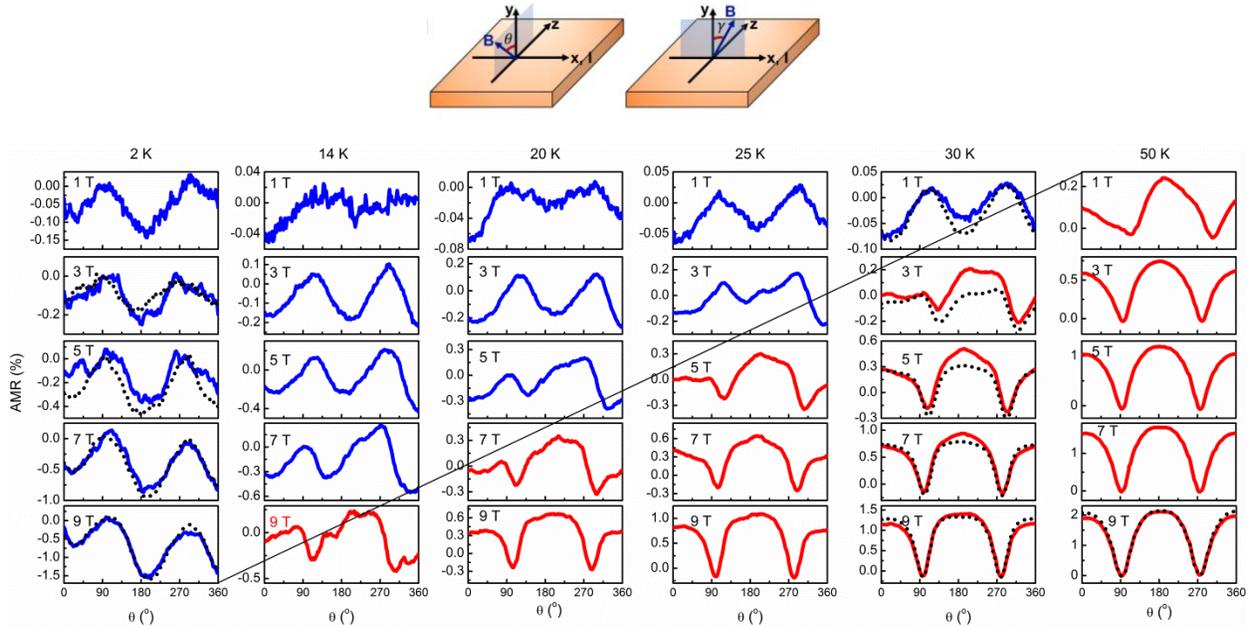

Figure S4: AMR ratio for (MI84)$_5$ highlighting the dynamics of the pahse shift and the modulation of the magnetic anisotropy as a function of temperature and magnetic field.

Detailed change in the AMR Oscillations for (MI84)$_5$ shows fine tuning of phase shift and sign reversal as function of temperature and magnetic field. The modulation in the magnetic anisotropy starts to appear only at 14 K for the application of 9 T magnetic field. The transition to in-plane magnetic easy axis systematically moves towards higher magnetic field with the increase in the temperature. For the lower temperature of 2 K and 10 K, the AMR ratio is negative with no magnetic field induced modulation in the anisotropy. For the higher temperature of 50 K, the modulation of anisotropy with the easy axis along in-plane direction is stabilizes with the positive AMR ratio throughout the range of the magnetic field of the measurements.

The similar treads in terms of magnitude, phase and sign of AMR is observed for γ geometry, presented by the dashed lines in the figure S of supplemental materials indicates that both the axis are isostructural in nature.

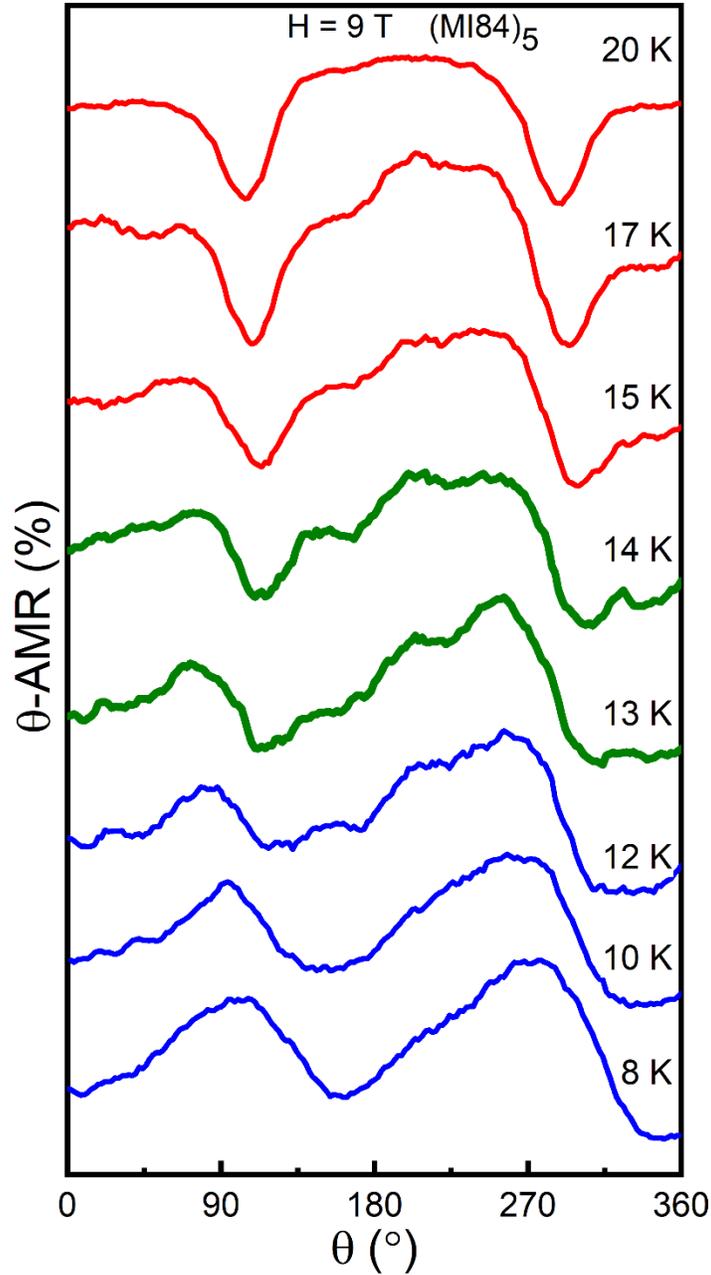

Figure S5: Evolution of the magnetic anisotropy in (MI84)$_5$ in the close interval of 8-20 K at H = 9 T.

The AMR measurements taken in the close interval of 8-20 K at H = 9 T indicates the tailoring of magnetic anisotropy via the rotation of the magnetization easy axis is very slow and gradual. Here, as seen from the figure x, a small shift in the phase of the AMR oscillations at 8 and 10 K observed however, the difference in the scattering amplitude for [100] and [$\bar{1}$00] direction suggests the destruction in the magnetic easy axis. Evident sign and phase reversal in AMR starts at 12 K and transit at 14 K which eventually completes at 20 K for the field strength of 9 T.

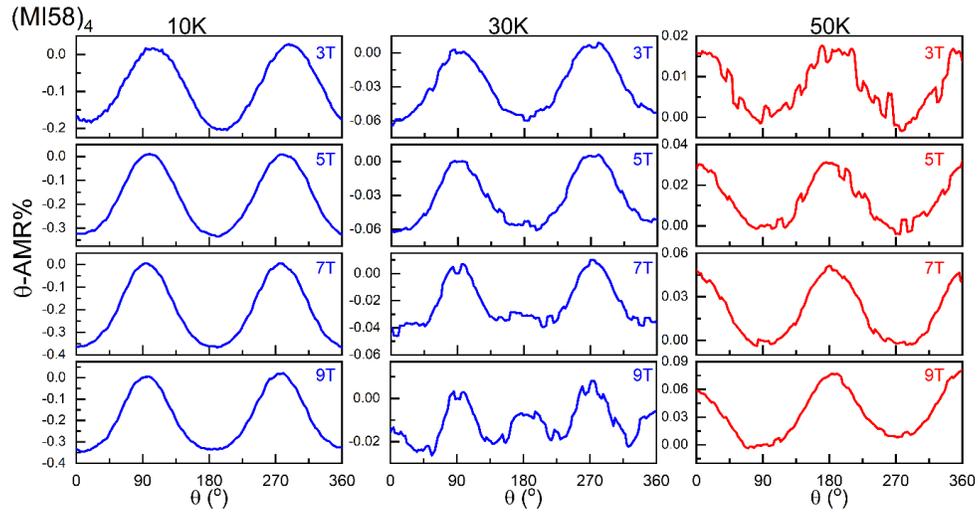

Figure S6: AMR oscillations with increasing magnetic field strength obtained at T = 10, 30 and 50 K for (MI58)$_4$ superlattice.

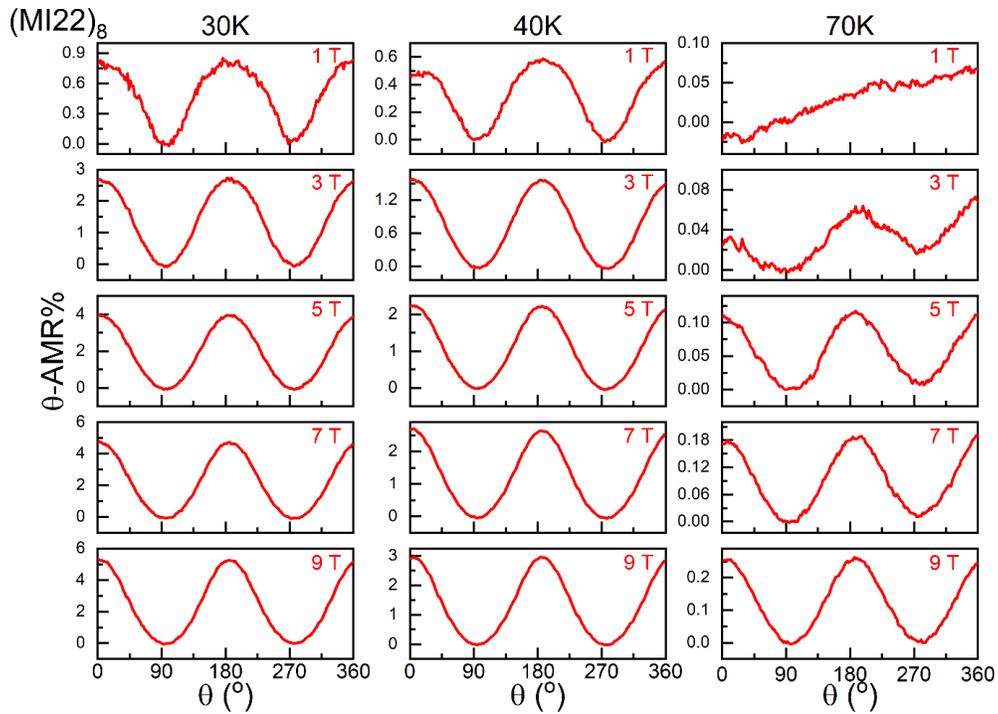

Figure S7: AMR oscillations with increasing magnetic field strength obtained at T = 30, 40 and 70 K for (MI58)$_4$ superlattice.

Figures S6 and S7 with the decrease in the strength of Rashba-SOC from (MI84)$_5$ to (MI58)$_4$, the transition accommodating phase shift moves towards higher temperature for H = 9 T whereas in (MI58)$_4$ no magnetic field induced such transition is observed throughout the temperature range of measurements. For (MI22)$_8$ with lowest strength of Rashba-SOC, this transition vanishes completely for the function of both temperature and magnetic field.

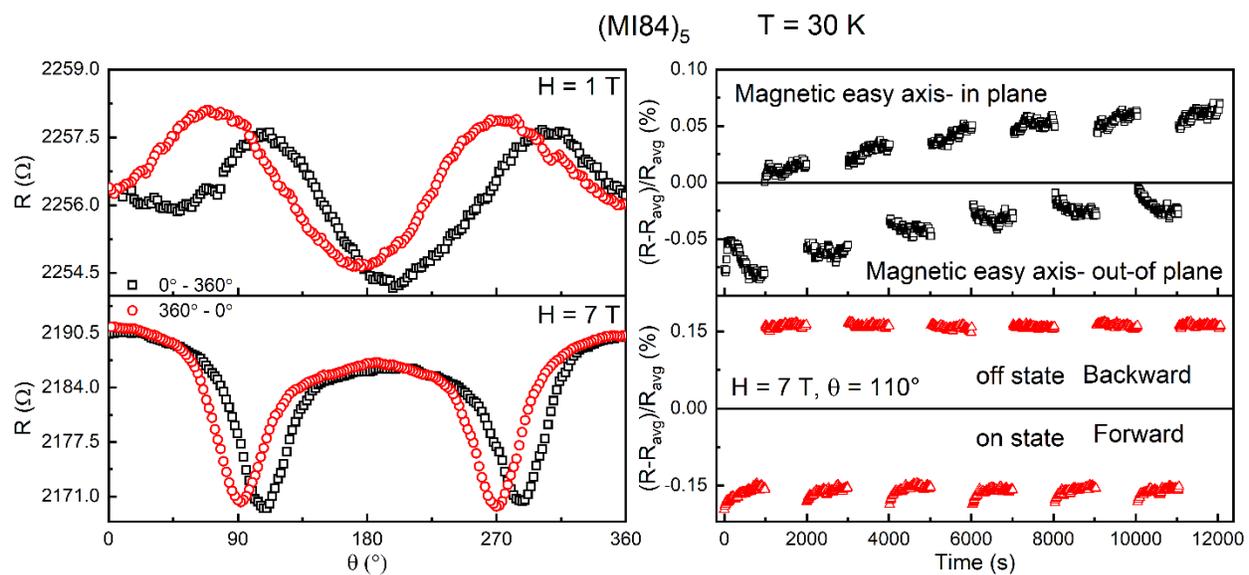

Figure S8: Angular dependence of magnetoresistance for (MI84)$_5$ measured in forward and reverse sweeping of position at T = 30 and H = 1 T and 7 T. Retention property of AMR obtained by switching between two resistance states at 110° obtained due to the hysteresis.

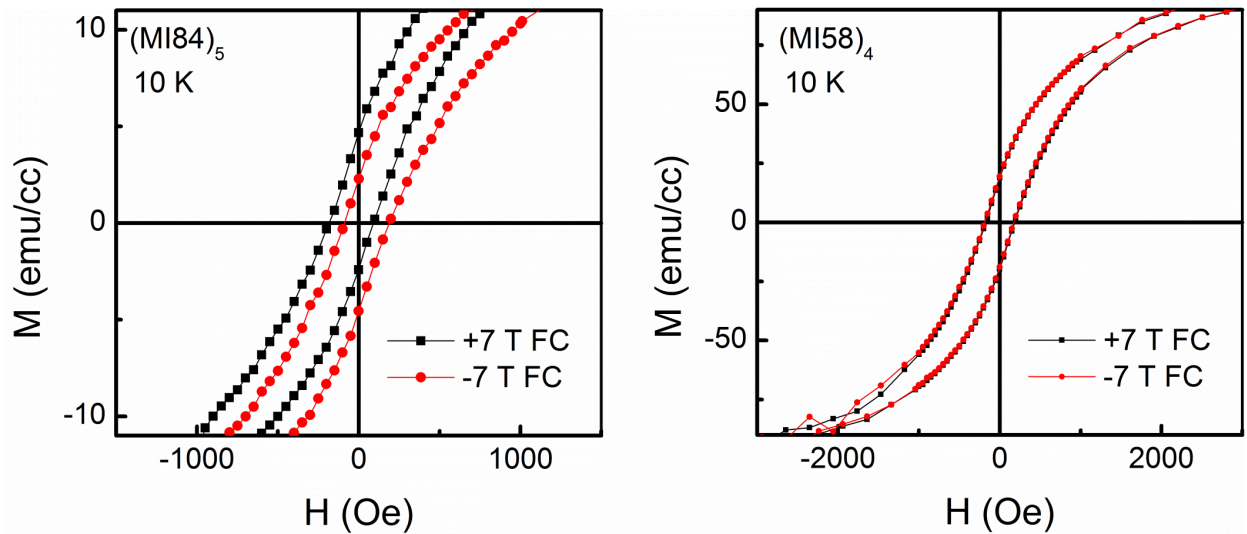

Figure S9: Exchange bias measurements obtained by measuring magnetization isotherms in +/- 7 T field cooling protocol at T = 10 K.

In $(MI84)_5$, asymmetrical M-H behaviour with shift in the opposite direction for +/- 7 T field cooled MH indicates the presence of AFM and FM phases giving rise to exchange bias field of ~ 35 Oe. On the other hand, no exchange bias field is observed for $(MI58)_4$ and $(MI22)_8$.

The charge-transfer across the interface decays exponentially from the interface to within the layer. As a result, a fraction of $Mn^{4+}$ proportional to charge transfer converts to $Mn^{3+}$ which induces a double-exchange governed largely canted AFM or a weak FM phase at the interface layer. The inner part $CaMnO_3$ layer remains weakly canted as it coincides with the exponential tail of charge transfer. The manifestation of $H_{EB}$ confirms the magnetic gradient across the interface.

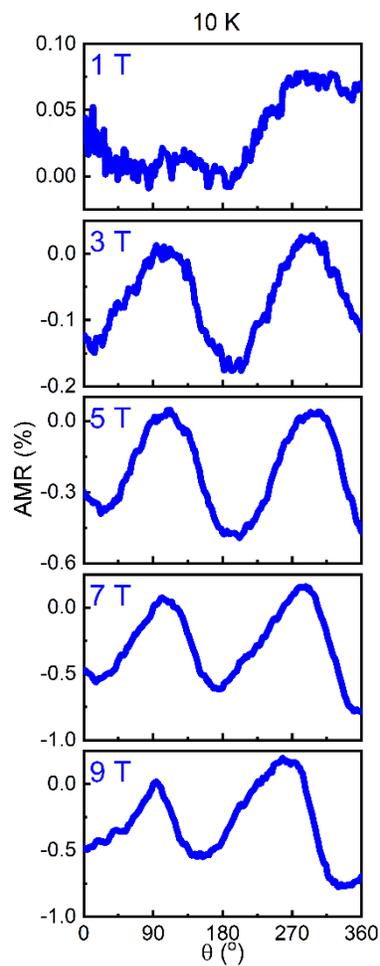

Figure S10: Magnetic field dependence of AMR ration of (MI84)$_5$ at 10 K.